\PassOptionsToPackage{unicode}{hyperref}
\PassOptionsToPackage{hyphens}{url}
\PassOptionsToPackage{dvipsnames,svgnames,x11names}{xcolor}
\documentclass[
  12pt]{article}

\usepackage{amsmath,amssymb,bm,subfigure}
\usepackage{adjustbox}
\usepackage{booktabs}
\usepackage{iftex}
\ifPDFTeX
  \usepackage[T1]{fontenc}
  \usepackage[utf8]{inputenc}
  \usepackage{textcomp} 
\else 
  \usepackage{unicode-math}
  \defaultfontfeatures{Scale=MatchLowercase}
  \defaultfontfeatures[\rmfamily]{Ligatures=TeX,Scale=1}
\fi
\usepackage{lmodern}
\ifPDFTeX\else  
\fi
\IfFileExists{upquote.sty}{\usepackage{upquote}}{}
\IfFileExists{microtype.sty}{
  \usepackage[]{microtype}
  \UseMicrotypeSet[protrusion]{basicmath} 
}{}
\makeatletter
\@ifundefined{KOMAClassName}{
  \IfFileExists{parskip.sty}{%
    \usepackage{parskip}
  }{
    \setlength{\parindent}{0pt}
    \setlength{\parskip}{6pt plus 2pt minus 1pt}}
}{
  \KOMAoptions{parskip=half}}
\makeatother
\usepackage{xcolor}
\makeatletter
\ifx\paragraph\undefined\else
  \let\oldparagraph\paragraph
  \renewcommand{\paragraph}{
    \@ifstar
      \xxxParagraphStar
      \xxxParagraphNoStar
  }
  \newcommand{\xxxParagraphStar}[1]{\oldparagraph*{#1}\mbox{}}
  \newcommand{\xxxParagraphNoStar}[1]{\oldparagraph{#1}\mbox{}}
\fi
\ifx\subparagraph\undefined\else
  \let\oldsubparagraph\subparagraph
  \renewcommand{\subparagraph}{
    \@ifstar
      \xxxSubParagraphStar
      \xxxSubParagraphNoStar
  }
  \newcommand{\xxxSubParagraphStar}[1]{\oldsubparagraph*{#1}\mbox{}}
  \newcommand{\xxxSubParagraphNoStar}[1]{\oldsubparagraph{#1}\mbox{}}
\fi
\makeatother

\usepackage{longtable,booktabs,array}
\usepackage{calc} 
\usepackage{etoolbox}
\makeatletter
\patchcmd\longtable{\par}{\if@noskipsec\mbox{}\fi\par}{}{}
\makeatother
\IfFileExists{footnotehyper.sty}{\usepackage{footnotehyper}}{\usepackage{footnote}}
\makesavenoteenv{longtable}
\usepackage{graphicx}
\makeatletter
\def\maxwidth{\ifdim\Gin@nat@width>\linewidth\linewidth\else\Gin@nat@width\fi}
\def\maxheight{\ifdim\Gin@nat@height>\textheight\textheight\else\Gin@nat@height\fi}
\makeatother
\setkeys{Gin}{width=\maxwidth,height=\maxheight,keepaspectratio}
\makeatletter
\def\fps@figure{htbp}
\makeatother

\makeatletter
\@ifpackageloaded{caption}{}{\usepackage{caption}}
\AtBeginDocument{%
\ifdefined\contentsname
  \renewcommand*\contentsname{Table of contents}
\else
  \newcommand\contentsname{Table of contents}
\fi
\ifdefined\listfigurename
  \renewcommand*\listfigurename{List of Figures}
\else
  \newcommand\listfigurename{List of Figures}
\fi
\ifdefined\listtablename
  \renewcommand*\listtablename{List of Tables}
\else
  \newcommand\listtablename{List of Tables}
\fi
\ifdefined\figurename
  \renewcommand*\figurename{Figure}
\else
  \newcommand\figurename{Figure}
\fi
\ifdefined\tablename
  \renewcommand*\tablename{Table}
\else
  \newcommand\tablename{Table}
\fi
}
\@ifpackageloaded{float}{}{\usepackage{float}}
\floatstyle{ruled}
\@ifundefined{c@chapter}{\newfloat{codelisting}{h}{lop}}{\newfloat{codelisting}{h}{lop}[chapter]}
\floatname{codelisting}{Listing}

\makeatother
\makeatletter
\@ifpackageloaded{caption}{}{\usepackage{caption}}
\@ifpackageloaded{subcaption}{}{\usepackage{subcaption}}
\makeatother

\ifLuaTeX
  \usepackage{selnolig}  
\fi
\usepackage[]{natbib}
\usepackage{bookmark}

\IfFileExists{xurl.sty}{\usepackage{xurl}}{} 
\hypersetup{
  pdftitle={Title},
  pdfauthor={Author 1; Author 2},
  pdfkeywords={3 to 6 keywords, that do not appear in the title},
  colorlinks=true,
  linkcolor={blue},
  filecolor={Maroon},
  citecolor={Blue},
  urlcolor={Blue},
  pdfcreator={LaTeX via pandoc}}

\newcommand{\anon}{1}

\begin{document}

\def\spacingset#1{\renewcommand{\baselinestretch}%
{#1}\small\normalsize} \spacingset{1}


\if1\anon  
{
  \title{\bf Electoral Polls and Economic Uncertainty: an Analysis of the Last Two  U.S. Presidential Elections}
  \author{Giampiero M. Gallo\hspace{.2cm}\\
  	New York University in Florence\\ CRENoS - Centro Ricerche Economiche Nord Sud\\
  	and \\
  	Demetrio Lacava\thanks{Corresponding author. Email: dlacava@unime.it}\hspace{.2cm}\\
Department of Economics, University of Messina\\
and \\
Edoardo Otranto\thanks{
	Edoardo Otranto gratefully acknowledges the \textit{Italian  PRIN 2022 grant (20223725WE) ``Methodological and computational issues in large-scale  series models for economics and finance".} } \\
	Department of Social Sciences and Economics, Sapienza University of Rome\\
 CRENoS - Centro Ricerche Economiche Nord Sud}
  \maketitle
} \fi

\if0\anon
{
  \bigskip
  \bigskip
  \bigskip
  \begin{center}
    {\LARGE\bf Title}
\end{center}
  \medskip
} \fi

\bigskip
\begin{abstract}
This paper examines the dynamic relationship between electoral polls and indicators of economic and financial uncertainty during the last two U.S. presidential elections (2020 and 2024). Using daily polling data on Donald Trump and measures such as the Aruoba-Diebold-Scotti Business Conditions Index, the 5-year Breakeven Inflation Rate, the Trade Policy Uncertainty index, and the VIX, we estimate conditional correlation models to capture time-varying interactions. The analysis reveals that in 2020, correlations between polls and uncertainty measures were highly dynamic and event-driven, reflecting the influence of exogenous shocks (COVID-19, oil price collapse) and political milestones (primaries, debates). In contrast, during the 2024 campaign, correlations remained close to zero, stable, and largely unresponsive to shocks, suggesting that entrenched polarization and non-economic events (e.g., assassination attempt, candidate changes) muted the economic channel. The study highlights how the interplay between voter sentiment, financial markets, and uncertainty varies across electoral contexts, offering a methodological contribution through the application of Dynamic Conditional Correlation models to political data and policy-relevant insights on the conditions under which economic fundamentals influence electoral dynamics.
\end{abstract}

\noindent%
{\it Keywords:} Financial market volatility; Inflation expectations; Business conditions; Trade policy; Dynamic Conditional Correlation; Political campaign analysis.
\vfill

\newpage
\spacingset{1.8} 

\section{Introduction}\label{sec-intro}

The U.S. presidential election polls are a valuable example of how political opinions evolve during an electoral campaign, offering important information to political analysts on changes in public sentiment over time. Public beliefs reflected in election polls may be influenced by both election-related events and exogenous shocks. By examining the evolution of potential voter preferences, it is possible to assess the extent to which exogenous events may shape the political outlook on what is deemed relevant during a campaign: these could be represented by economic fundamentals such as inflation or the state of the economy, financial market movements or specific issues related to trade policies which have gained in relevance in recent years. 

The transmission mechanisms between uncertainty measured by various economic and financial variables and voters’ political preferences are complex. On financial markets, the influence of the corporate actors on the U.S. political activity through legal contributions to a campaign (Political Action Committees -- PACs -- and Super PACs) aim at steering a candidate toward the support of policies that benefit them. That enforces a correlation between financial market behavior and the dynamics of electoral polls: for instance, if a candidate backed by a specific industry leads in the polls, it is plausible that stocks in that sector could experience price increases in anticipation of favorable policy outcomes. By the same token, new positions expressed by the candidates generate a reaction in the markets that may be perceived as an expert judgement by the larger public. In such a context, empirical research shows that financial institutions and large corporations shape policy debates and electoral outcomes through campaign contributions, lobbying, and public speeches \citep[see, for example,][]{Gilens:Page:2014}. Their actions in particular, can affect legislative priorities, regulatory frameworks, and trade policies, often aligning political decisions with the interests of capital markets. This would suggest that political consensus could be related to financial markets, on the one hand, and to economic and trade uncertainty, on the one other. As for the stock market, \citet{Abolghasemi:Stanko:2021} find a causal relationship going from the U.S. presidential election outcome to the dynamics in asset prices. However, it emerges that the stock markets response to new political information is not immediate, highlighting inefficiency in pricing this kind of information \citep{Flynn:Tarkom:2025}. Furthermore, political information is found to be linked to stock volatility as well \citep{Herold:Kanz:Muck:2021, Hartwell:Hubschmid_Vierheilig:2024}. 

From a different perspective, a large body of political economy and electoral studies emphasizes the role of economic fundamentals in shaping voter sentiment and electoral support. One of the most studied variables is inflation, which is seen as proxies for the economic environment that voters directly experience. For example, \citet{Palmer:Whitten:1999} shows that voters are primarily concerned with unexpected inflation; similarly, \citet{Doti:Campbell:2024} find that inflation is a key determinant of presidential election outcomes over long horizons, while \citet{Binder:Kamdar:Ryngaert:2024} show that perceptions of inflation are strongly politicized, with partisan divides shaping how voters interpret inflation dynamics. More recently, \citet{Baccini:Weymouth:2025} show that individuals who are more affected by inflation tend to favor Republican candidates.

To the best of our knowledge,  the branch of literature which investigates the link between election outcome and trade uncertainty is still narrow. A notable exception is a recent analysis by \citet{Handley:Limao:2022} highlights how electoral contests and regime uncertainty are natural drivers of trade  uncertainty, as measured through the Trade Policy Uncertainty (TPU) indices developed by \citet{Caldara:Iacoviello:Molligo:Prestipino:Raffo:2020}. In particular they show how TPU indicators spike around episodes of political change and policy discontinuity. 

\begin{figure}[t]
	\centering
	{\includegraphics[height=9cm,width=16cm]{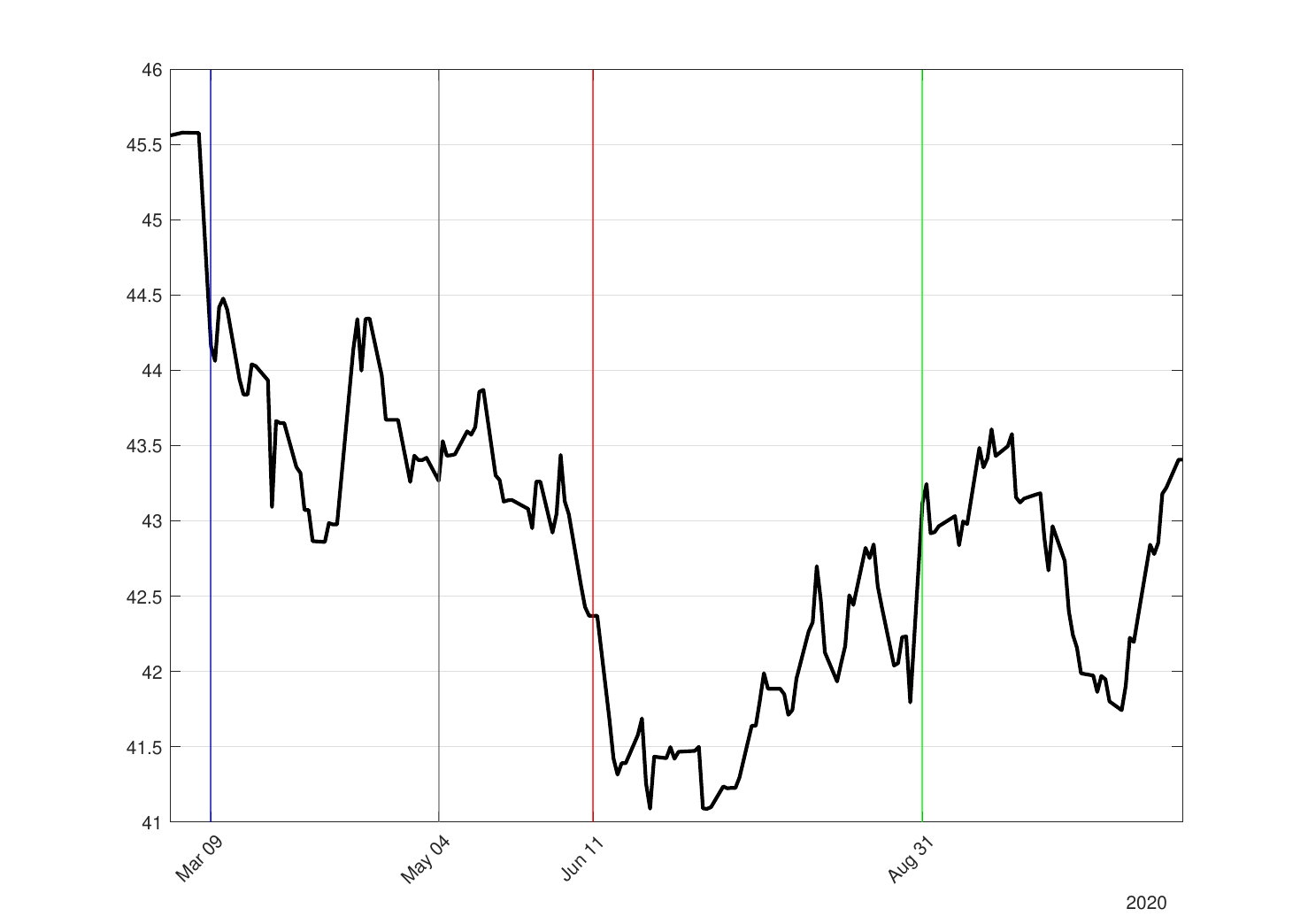}}
	\caption{Time-series of the polls in favor of Trump during the 2020 electoral campaign. The vertical lines represent some events related to inflation (blue vertical line), business condition (gray), VIX (red), and trade uncertainty (green).  Sample period: March 5, 2020 -  November 3, 2020.}
	\label{fig:polls}
\end{figure}

To give an example of what comes ahead, in the evolution of Trump's polling support during 2020 (shown in Figure \ref{fig:polls}), we marked with vertical lines days when some variables taken to represent major economic and financial conditions signal relevant occurrences. Note that the reduction in inflation on March 9 (blue line) coincides with the beginning of a decline in Trump's support. Conversely, a gradual increase in Trump's polls follows the improvement in business conditions recorded on May 4 (gray line). A sharp reduction is then observed in June, particularly after the market crash on June 11, when the VIX index jumped up by 48\% in one day (from 27.6 to 40.8). Interestingly, at the end of August, Trump's polls began to recover slightly, in correspondence with the surge of the U.S.–China tensions and global trade disruptions well captured by the Trade Policy Uncertainty Index on August 31 (+133\% green line).

This paper aims at assessing how several sources of uncertainty (i.e., financial, economic, and trade uncertainty) influence electoral polls by estimating some time-varying correlation models across the selected time series. We refer to the Dynamic Conditional Correlation \citep[DCC,]{Engle:2002b} model, where correlations evolve over time according to GARCH-type features \citep{Bollerslev:1986}. For relevant variables, we take the VIX index as a representative measure of stock markets uncertainty. Economic uncertainty is represented here by a daily series measuring expected inflation (extracted from bond rates indexed to inflation) and the Aruoba-Diebold-Scotti Business Conditions Index \citep[ADS,][]{Aruoba:Diebold:Scotti:2009}. As it is well known, unstable inflation generates uncertainty by reducing purchasing power in households and increasing costs in businesses, by shaping central banks actions which may increase financial burdens and discourage investments. 
The ADS daily index allows for insights into current and future economic trends, tracking how the information about new economic data releases can be filtered to represent real business conditions. Similarly to the Economic Policy Uncertainty index \citep[EPU,][]{Baker:Bloom:Davis:2016}, the TPU index is constructed by counting the frequency of joint occurrences of ``trade policy'' and ``uncertainty'' terms in major newspapers. The variable of interest is the U.S. presidential election daily national polls for candidate Trump, in the two most recent general elections of 2020 and 2024.


The empirical application reveals interesting results. In particular, the two election campaigns show very different features as characterized by the estimated correlations with uncertainty: a clear and time-varying relationship in 2020, which appears to vanish in 2024. We take this to be a sign of a fundamental difference in the communication strategy with polarization of opinions and a diffusion of effective messages in the later campaign that did not necessarily reflect the actual state of the economy.

The paper is structured as follows. In the next section, we describe the conditional correlation models considered. Section \ref{sec:descriptive_analysis} is devoted to a preliminary investigation of the relationship between polls and uncertainty variables, with variable cleaning, while Section \ref{sec:results} discusses the correlation dynamics underlying the divergent behaviors observed in the two election periods.
 Some final remarks conclude the paper.
 
\section{Dynamic Correlation Models}\label{sec:model}
From a computational perspective, modeling covariances and correlations poses two main challenges \citep{Bauwens:Laurent:Rombouts:2006}: (i) ensuring that the estimated covariance (or correlation) matrix is positive definite, and (ii) addressing the curse of dimensionality, whereby the number of parameters grows exponentially with the number of series, potentially making the model computationally intractable. A parsimonious approach to these issues is provided by the Dynamic Conditional Correlation (DCC) model \citep{Engle:2002b}. In this framework, \citet{Engle:2002b} proposes a two-step estimation procedure: in the first step, conditional variances are estimated, and in the second step, standardized (de-GARCHed) residuals are used to estimate conditional correlations.

Let $\bm{y}_t$ be an ($N \times 1$) vector 
of residuals from an AutoRegressive Moving Average (ARMA) model,  with conditional covariance matrix $\mathbf{H}_t$. Specifically, we assume  
$$\bm{y}_t| F_{t-1} \sim N(0, \mathbf{H_t}), \qquad t = 1, \ldots , T$$ 
and
$$\mathbf{H_t} = \mathbf{S_t R_t S_t},$$
where $F_{t-1}$ denotes the information set, $\mathbf{S}_t$ is a diagonal matrix of conditional standard deviations, and $\mathbf{R}_t$ is the conditional correlation matrix.

As proposed by \citet{Engle:2002b}, for each series \textit{i}, conditional standard deviations are obtained from a univariate GARCH-type model \citep{Bollerslev:1986}, where the conditional variance $h^2_{i,t}$ (the $i$-th diagonal element of $\mathbf{H_t}$)  is defined as
\begin{equation}
	h^2_{i,t}=\omega_i+\alpha_i y^2_{i,t-1}+\beta_i h^2_{i,t-1}, \qquad i=1,\dots,N \label{eq:garch}
\end{equation}
Here $ y^2_{i,t-1}$ captures the most recent shock, and $h^2_{i,t-1}$ reflects the impact of the lagged conditional volatility. Stationarity requires $(\alpha_i+\beta_i)<1$, while $\omega_i>0,\alpha_i\ge 0,\beta_i\ge 0$ ensure positivity.

After estimating the univariate GARCH models, the de-GARCHed residuals are computed as 
\begin{equation}
\tilde{\epsilon}_{i,t} = y_{i,t}/h_{i,t}.\label{eq:degarched_res}
\end{equation} 

The second-step estimation then provides $\mathbf{R}_t$, using a dynamic specification. In particular, the model is defined as:

\begin{align}
	\mathbf{R}_t =& \tilde{\mathbf{Q}}_t^{-1}\mathbf{Q}_t\tilde{\mathbf{Q}}_t^{-1}, 
	\quad \tilde{\mathbf{Q}}_t = \mathrm{diag}(\sqrt{q_{11,t}}, \ldots, \sqrt{q_{nn,t}}),\notag  \\[2mm]
	\mathbf{Q}_t = &(1-a-b)\bar{\mathbf{R}} + a\tilde{\mathbf{Q}}_{t-1}\tilde{\bm{\epsilon}}_{t-1}\tilde{\bm{\epsilon}}_{t-1}^\prime\tilde{\mathbf{Q}}_{t-1}  
	+ b\mathbf{Q}_{t-1}\label{eq:dcc}
\end{align}
Here, $\bar{\mathbf{R}}$ denotes the unconditional correlation matrix, estimated  from the sample correlation of $\tilde{\mathbf{Q}}_{t}\tilde{\bm{\epsilon}}_{t}$. As noted by \citet{Aielli:2013}, consistency is ensured by allowing the second term to depend on the lagged outer product of the de-GARCHed residuals. This term, with coefficient $a$, captures the impact of the most recent shocks on correlations, while the autoregressive nature of correlations is represented by coefficient $b$. Model stationarity requires $a+b<1$, while $a,b>0$ ensure positive definiteness of $\bm{Q}_t$.

Under the hypothesis of conditional normality, estimation is performed by Maximum Likelihood, with log-likelihood available in closed form and expressed as
\begin{equation}
	L(\bm{\theta}) = -\frac{TN}{2}log(2\pi) -\frac{T}{2} \log\left( |\mathbf{R_t}| \right) 
	-\frac{1}{2} \sum_{t=1}^T 
	\tilde{\bm{\epsilon}}_{t,\cdot} \, 
	\mathbf{R_t}^{-1} \, 
	\tilde{\bm{\epsilon}}_{t,\cdot}^\prime, \label{eq:loglik}
\end{equation}
where $\bm{\theta}$ denotes the vector of unknown parameters and $\tilde{\bm{\epsilon}}_{t,\cdot}$ the $t$-th row of the matrix of de-GARCHed residuals. 

Although the DCC combines parsimony and computational efficiency with the ability to capture dynamic correlations even for a large set of series, it imposes the same $a$ and $b$ coefficients for each pair of series. Furthermore, in this framework the coefficients are not allowed to vary over time, limiting the possibility of fully exploiting the available information by accounting for the specific impact of different types of news. In this regard, \citet{Bauwens:Otranto:2020} propose the NonLinear AutoRegressive Correlation (NARLC), which
extends the DCC by considering a time-varying coefficient $a_t$ in Eq. \ref{eq:dcc}. The difference with respect to the standard DCC is in the $\bm{Q}_t$ dynamics, which is specified as

\begin{align}
	\mathbf{Q}_t = &(1-a-b)\bar{\mathbf{R}} + a\bm{A_t}\odot\tilde{\mathbf{Q}}_{t-1}\tilde{\bm{\epsilon}}_{t-1}\tilde{\bm{\epsilon}}_{t-1}^\prime\tilde{\mathbf{Q}}_{t-1}  
	+ b\mathbf{Q}_{t-1}
\end{align}
where $\odot$ is the Hadamard (element by element) product and
\begin{align}
	\bm{A_t}=& exp^\odot\left[\phi_A(\bm{R}_{t-1}-\bm{J}_n) \right]=\left[exp^\odot\left(\phi_A\bm{R}_{t-1}\right)\right]/exp(\phi_A),\quad \phi_A\geq 0.	\label{eq:nlarc}
\end{align}
Here, $\bm{J}_n$ is a $n\times n$ matrix of ones. The use of $exp^\odot$ (i.e., the Hadamard exponential operator) together with $\phi_A\geq0$ ensures the positiveness of $\bm{A}_t$, thereby preserving the positive definiteness of the correlation matrix.
 The main advantage of this specification is that $\bm{A}_t$ is not only time-varying but also specific for each pair of series, as measured by each entry $a_{ij,t}$. Within this setting, the model allows for specific dynamics for each correlation, while introducing only one additional parameter, $\phi_A$, compared with the standard DCC. In detail, this specification considers two separate autoregressive dependencies. The first one is the linear dependence on the
lagged covariance $q_{ij,t-1}$, with the scalar parameter $b$, as in the scalar DCC model; the other one, included in the matrix $\bm{A}_t$, adds a nonlinear dependence on the lagged conditional correlation \citep[see][for further details]{Bauwens:Otranto:2020}. As a result, the NLARC nests the standard DCC, which is obtained by imposing $\phi_A=0$.

As a benchmark, we consider the Constant Conditional Correlation (CCC) model of \citet{Bollerslev:1990}, where the correlation matrix $\mathbf{R}$ is assumed to be time-invariant and estimated by the sample correlation of the de-GARCHed residuals. In this framework, covariances remain time-varying because they scale with the conditional volatilities obtained from the first-step GARCH estimates. We refer to \citet{Engle:Kroner:1995} for the conditions needed to ensure the positive definiteness of $\mathbf{H}_t$.

\section{Preliminary Analysis and First-step Estimation}\label{sec:descriptive_analysis}

To provide some details, the analysis is based on the polls in favor of Trump for the 2020 and 2024 electoral campaigns, as resulting from a daily survey conducted by FiveThirtyEight.\footnote{Data are available at \url{https://github.com/fivethirtyeight/data/tree/master/polls}. Data are also available for individual US states, which also show interesting spatio-temporal dynamics \citep[see][]{Gallo:Lacava:Otranto:2025}.} As regards uncertainty, we consider different proxies measuring different sources of uncertainty. First, we consider the Aruoba-Diebold-Scotti Business Condition index \citep[ADS,][]{Aruoba:Diebold:Scotti:2009}, which tracks real business conditions at a daily frequency,\footnote{The dataset, provided by the Federal Reserve Bank of Philadelphia, are available at \url{https://www.philadelphiafed.org/surveys-and-data/real-time-data-research/ads}.} basing on seasonally adjusted economic components.\footnote{These components: the weekly Initial Unemployment Claims; the monthly Payroll Employment; the monthly Industrial Production; the monthly Real Personal Income Less Transfer Payments; the monthly Real Manufacturing and Trade Sales and; the quarterly Real GDP.} Second, we consider the 5-Year Breakeven Inflation Rate.\footnote{The data series can be retrieved from \url{https://fred.stlouisfed.org/series/T5YIE}.} It is a market-based measure of expected inflation, calculated as the difference between the yield on a 5-Year Treasury Constant Maturity Security (nominal yield) and the yield on a 5-Year Treasury Inflation-Indexed Constant Maturity Security (real yield). Third, we consider the Trade Policy Uncertainty index \citep[TPU][]{Caldara:Iacoviello:Molligo:Prestipino:Raffo:2020}, which -- similarly to the Economic Policy Uncertainty \citep[EPU][]{Baker:Bloom:Davis:2016} -- is constructed by counting the share of newspaper articles in major U.S. outlets that contain a set of keywords related to both uncertainty and trade policy.\footnote{The series is accessible at \url{https://www.matteoiacoviello.com/tpu.htm}.} Finally we consider the Chicago Board Options Exchange (CBOE) volatility index, VIX,  a forward-looking measure of financial uncertainty, reflecting the market's expected 30-day volatility of the S\&P 500 index derived from option prices, as provided by Yahoo Finance. We consider two distinct electoral campaign periods: from March 5 to November 3, 2020, covering the presidential election campaign of that year, and from March 8 to November 5, 2024, corresponding to the 2024 general election. We focus on polls in favor of Donald Trump, who was the Republican candidate in both races, to enable a direct comparison between the two periods.

In the empirical application, the variables are considered in logarithms, with the exception of ADS and TPU, which contain zeros and negative values. After testing for stationarity using the Augmented Dickey-Fuller \citep[ADF,][]{Dickey:Fuller:1979} test, we conduct a preliminary analysis to identify the ARMA data-generating process (DGP) of each series and to estimate the GARCH model, as required in the two-step estimation of DCC-type models. At the 5\% significance level of the ADF test, the only stationary series are ADS in 2020 and 2024, and TPU in 2024. Accordingly, we consider these variables in levels, while taking first differences for the remaining series.

Table \ref{tab:res_preliminary} reports the identified DGP for each series, where $\Delta$ denotes the use of first differences for non-stationary series. For the 2020 period, all series, except TPU, exhibit signs of autocorrelation in the squared residuals, as indicated by the rejection of the ARCH-LM test for residual heteroskedasticity \citep{Engle:1982}. Conversely, in 2024, heteroskedasticity is detected only for ADS and VIX. Based on these results, we estimate a GARCH(1,1) model for the series affected by residual heteroskedasticity, thereby obtaining the de-GARCHed residuals $\tilde{\epsilon}$, as in Eq. \ref{eq:degarched_res}; for the remaining series, $\tilde{\epsilon}$ is computed using the unconditional standard deviation.

\begin{table}[t]
	\centering
	\footnotesize
	\caption{Data-generating process and p-value of the ARCH-LM \citep{Engle:1982} test for residual heteroskedasticity.}\label{tab:res_preliminary}
		\begin{tabular}{lccclcc}
			\toprule
			&\multicolumn{2}{c}{2020} & &&\multicolumn{2}{c}{2024}\\
			& DGP  & p-value ARCH test & & &DGP  & p-value ARCH test   \\
$\Delta polls$ & ARMA(0,0) & 0.907 & & $\Delta polls$ & ARMA(1,1) & 0.158\\
$ADS$ & ARMA(2,1) & 0.000 & & $ADS$ & ARMA(3,3) & 0.000\\
$\Delta inflation$ & ARMA(3,1) & 0.000 & & $\Delta inflation$ & ARMA(0,0) & 0.832\\
$\Delta TPU$ & ARMA(0,1) & 0.927 & & $TPU$ & ARMA(1,0) & 0.330\\
$\Delta VIX$ & ARMA(1,0) & 0.029 & & $\Delta VIX$ & ARMA(0,0) & 0.000\\
		\bottomrule
		\end{tabular}
\end{table}

\begin{figure}[t]
	\centering
	\subfigure[vs ADS]{\includegraphics[height=4.5cm,width=6.7cm]{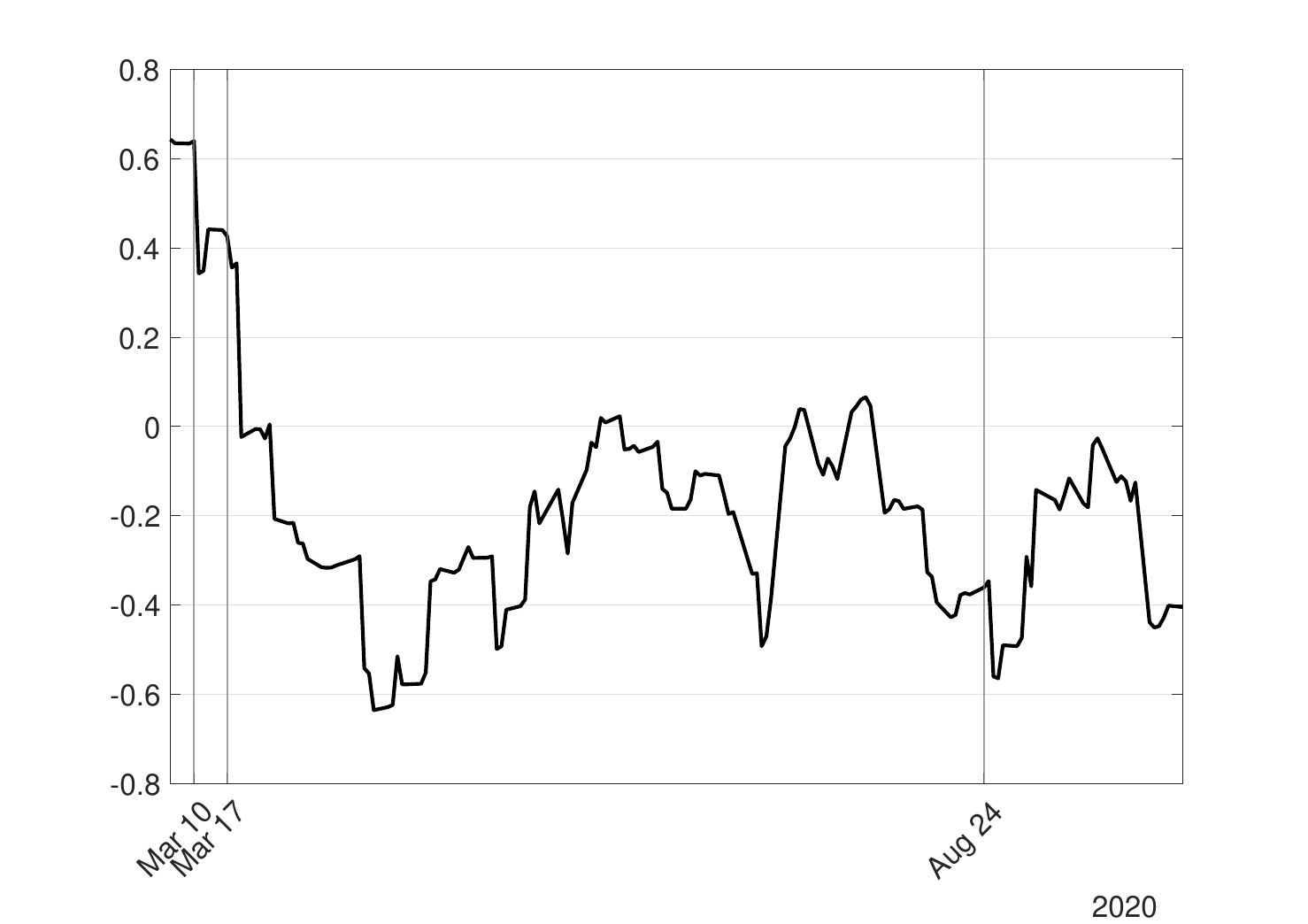}}
	\subfigure[vs inflation]{\includegraphics[height=4.5cm,width=6.7cm]{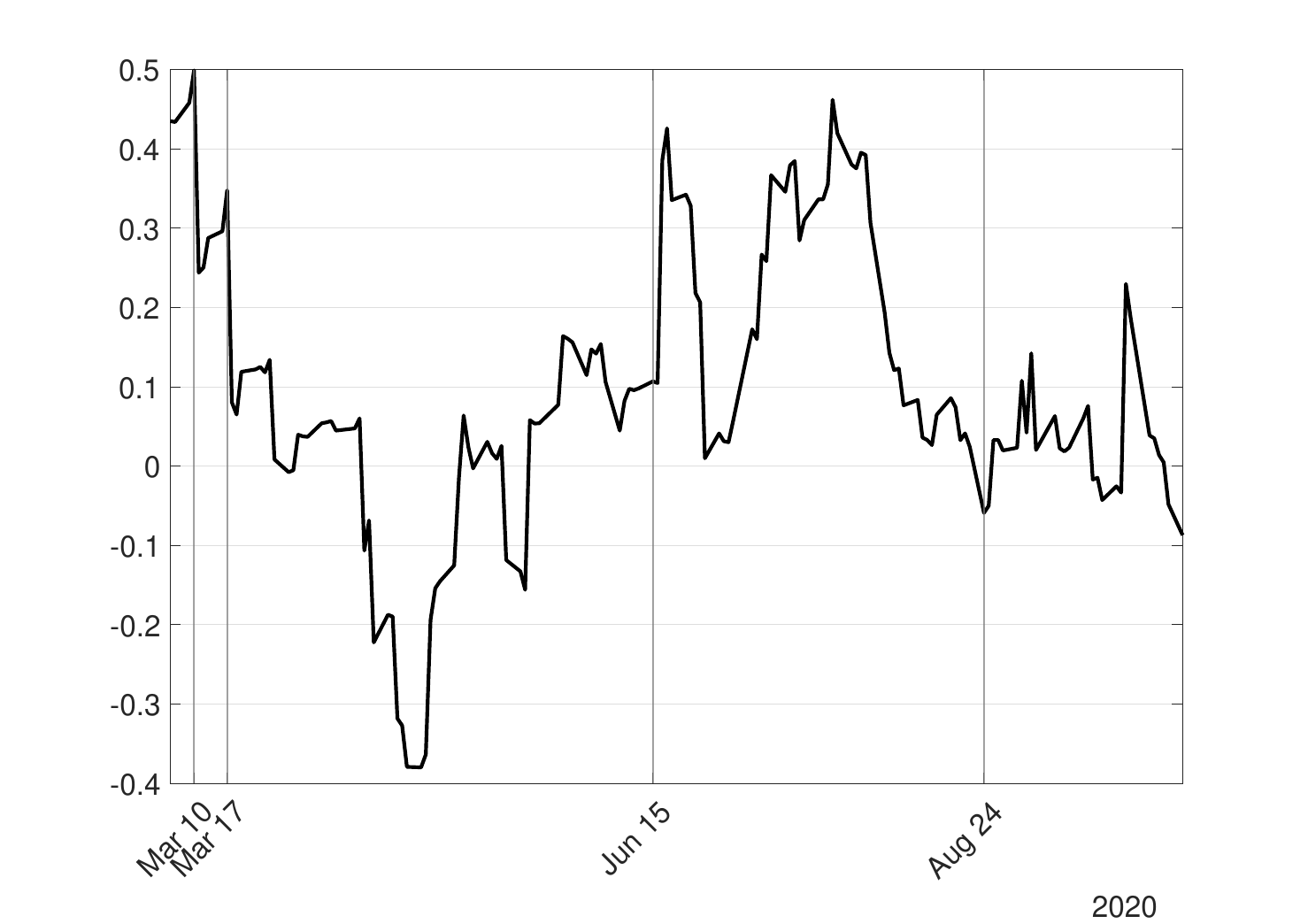}}\\
	\subfigure[vs TPU]{\includegraphics[height=4.5cm,width=6.7cm]{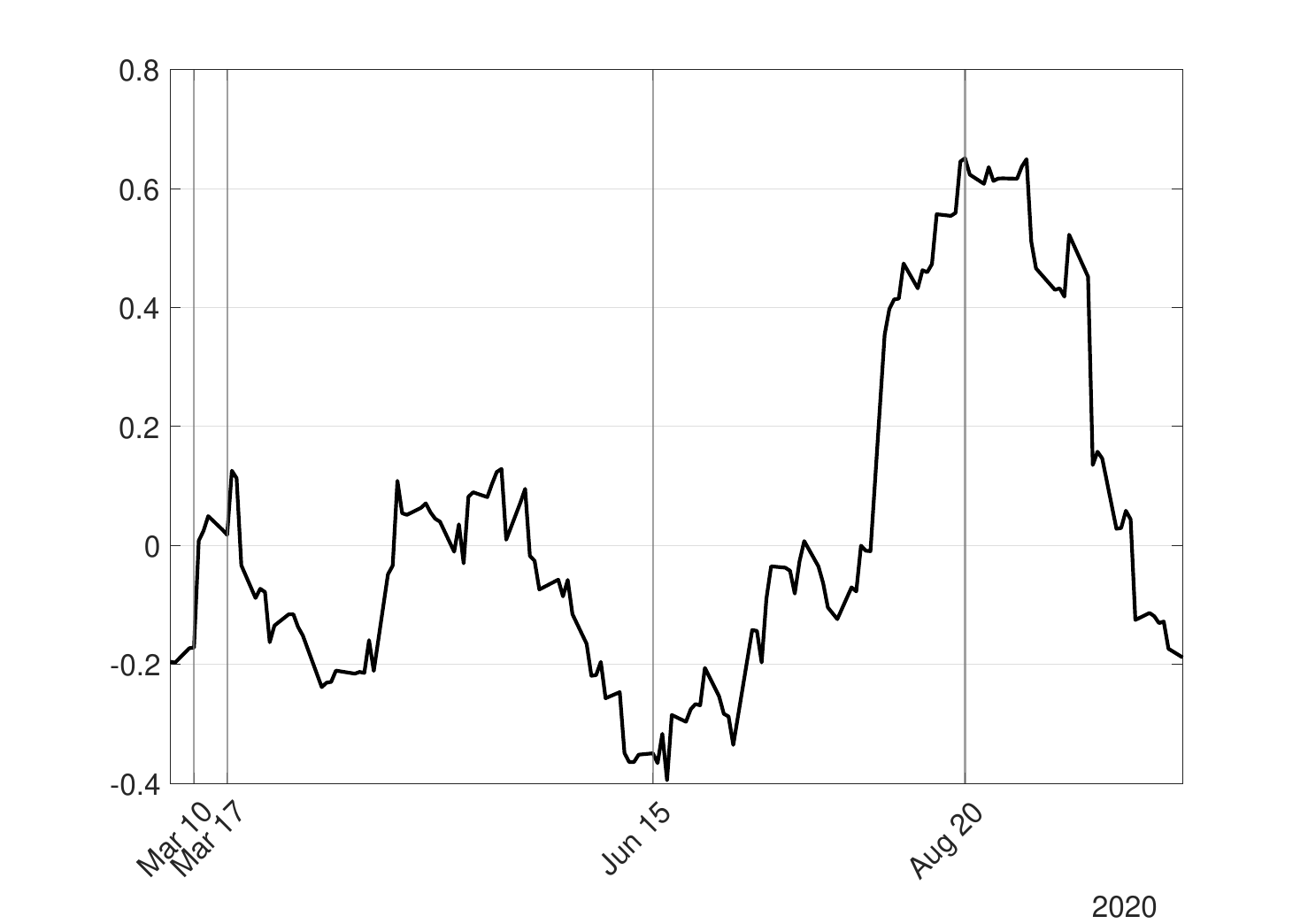}}
	\subfigure[vs VIX]{\includegraphics[height=4.5cm,width=6.7cm]{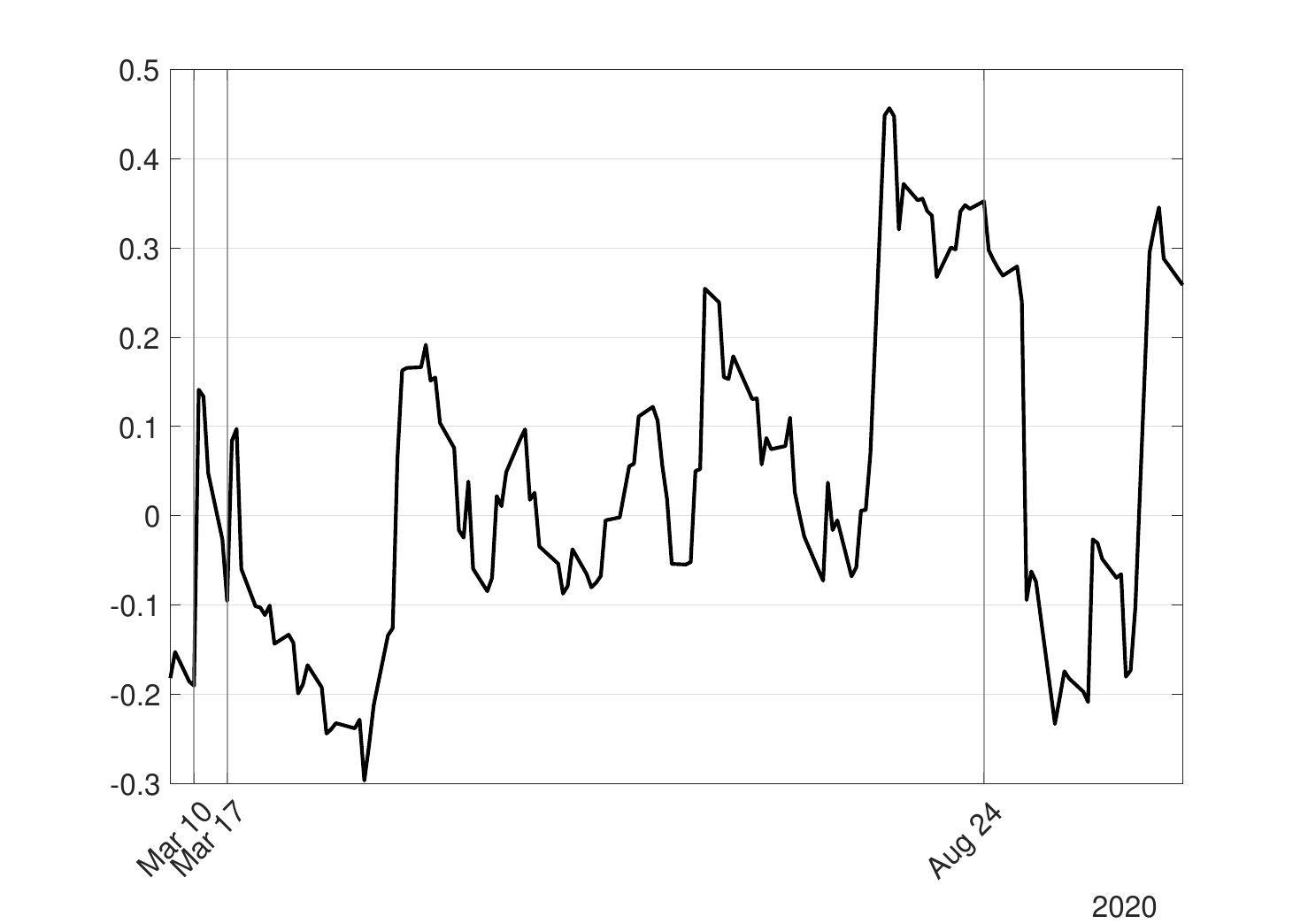}}
	\caption{5-day rolling correlations (black line) end relevant electoral events (gray vertical lines, see text). Sample period: March 5, 2020 -  November 3, 2020.}
	\label{fig:rolling_2020}
\end{figure}

As a check for the suitability of conditional correlation models, taking the 2020 period as an example, Figure \ref{fig:rolling_2020} displays the evolution of the 5-day rolling correlations (black line) between the de-GARCHed poll series and the de-GARCHed sources of uncertainty for the 2020 period. The vertical lines mark key electoral events: the primary election days (March 10 and March 17), the postponement of the Republican convention (June 15), the conclusion of the Democratic convention (August 20), and the beginning of the Republican convention (August 24).

All rolling correlations are clearly time-varying, frequently switching from negative to positive values around major events, which supports the use of DCC-type models. For example, correlations with the ADS business conditions index (Panel a) fluctuate substantially without a persistent pattern, showing a marked decline after the March 17 primaries, while increasing again at the start of the Republican convention in August. Correlations with inflation (Panel b) tend to remain positive for extended periods, particularly around mid-July and late August -- coinciding with the party conventions -- suggesting that higher inflation expectations were often associated with stronger support for Trump. Correlations with TPU (Panel c) alternate between positive and negative values, reflecting the episodic nature of trade policy issues in the 2020 campaign. Finally, correlations with the VIX (Panel d) are highly volatile and shift signs rapidly, becoming negative after the identified events, indicating that the relationship between financial uncertainty and Trump's polling support changed direction rather than maintaining a stable sign. Overall, this evidence supports the view that correlations evolve dynamically, as further emphasized in the next section with the estimation results.

\section{Second-step Estimation: Dynamic and Constant Correlations}\label{sec:results}
The top panel of Table \ref{tab:results} reports the estimation results for the DCC and NLARC models. In the CCC model, the estimated correlations are not explicitly reported, since the model relies on the sample correlation of the standardized (de-GARCHed) residuals \citep{Bollerslev:1990}. 

\begin{figure}[t]
	\centering
	\subfigure[vs ADS]{\includegraphics[height=4.5cm,width=6.7cm]{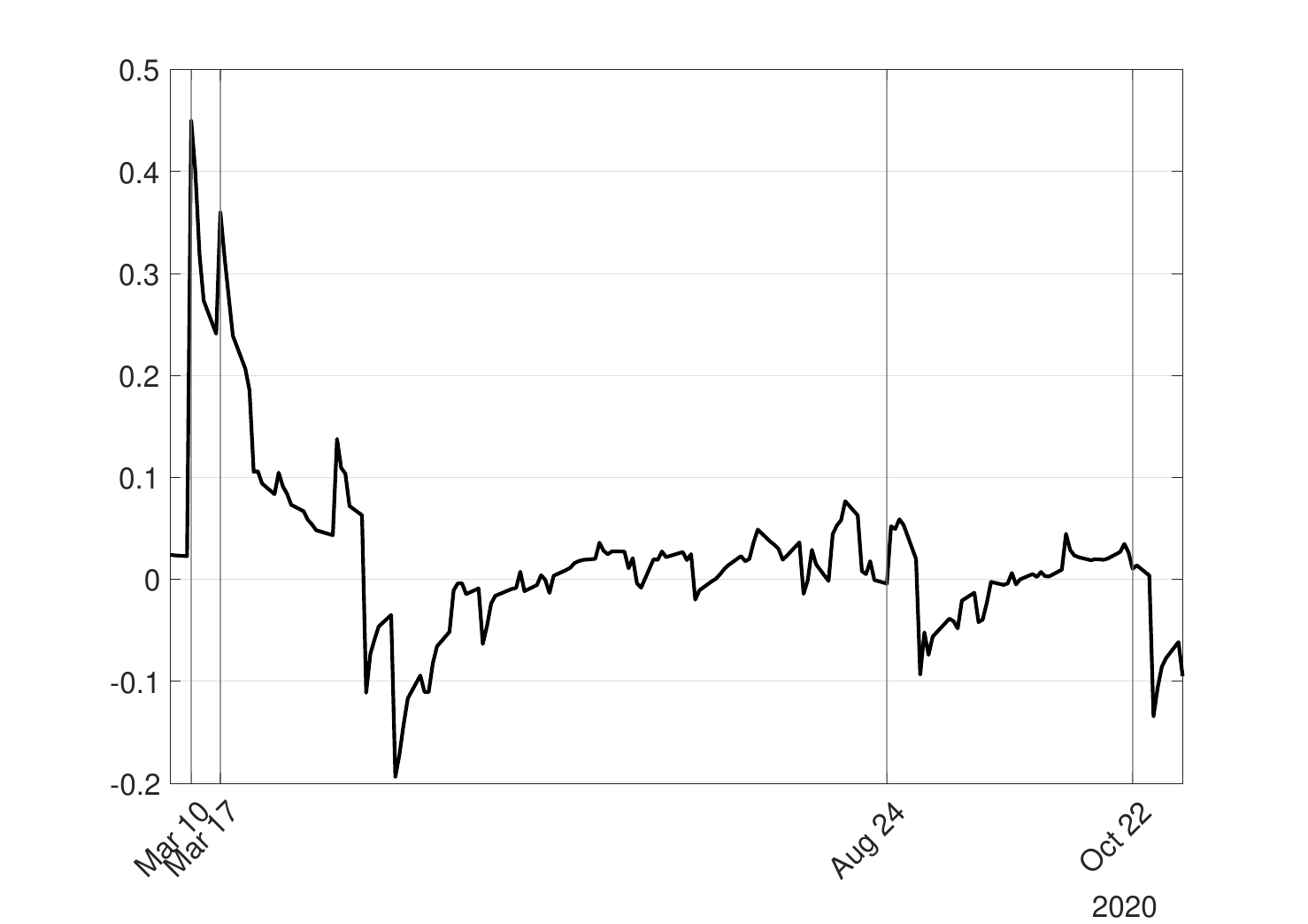}}
	\subfigure[vs inflation]{\includegraphics[height=4.5cm,width=6.7cm]{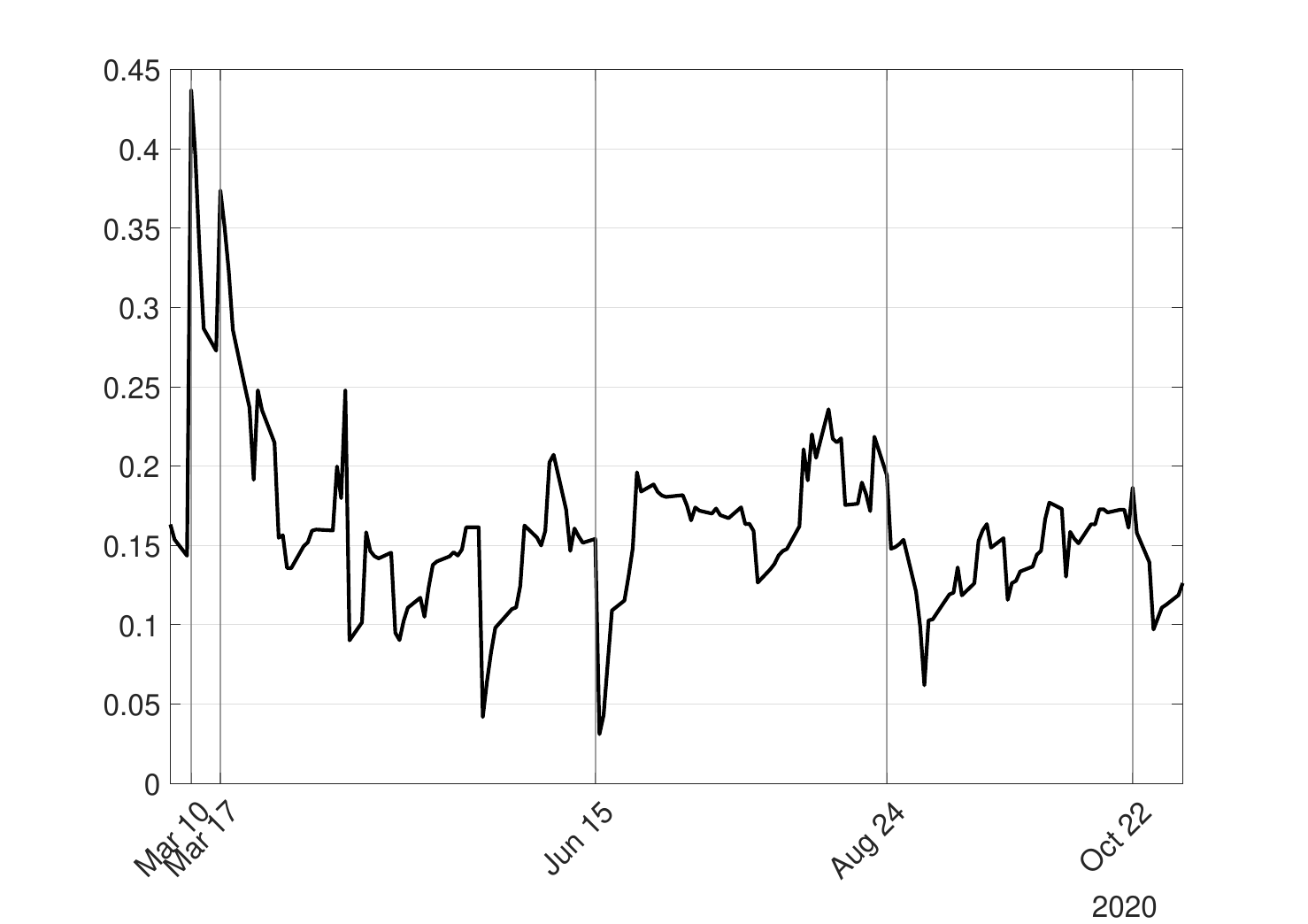}}\\
		\subfigure[vs TPU]{\includegraphics[height=4.5cm,width=6.7cm]{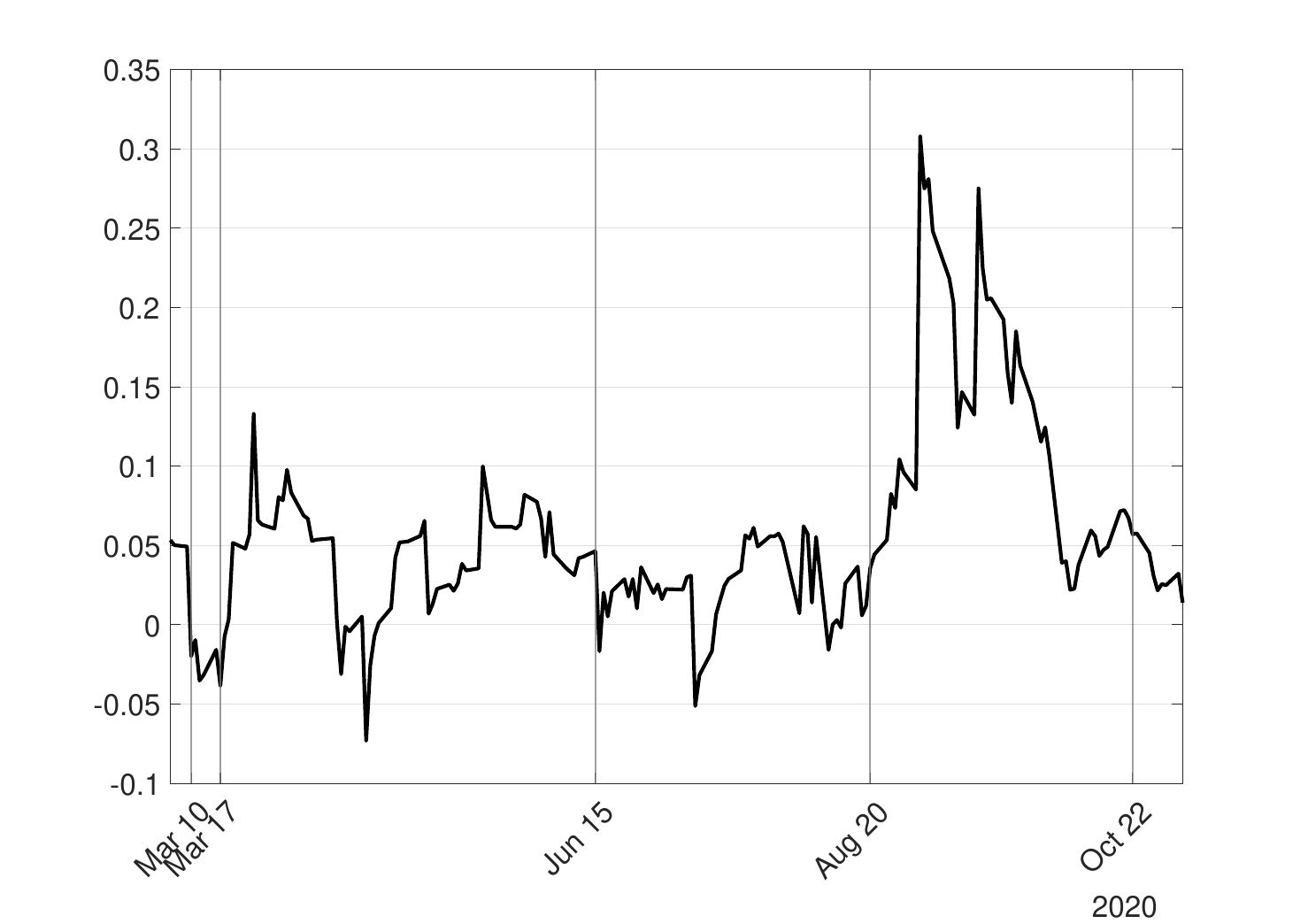}}
	\subfigure[vs VIX]{\includegraphics[height=4.5cm,width=6.7cm]{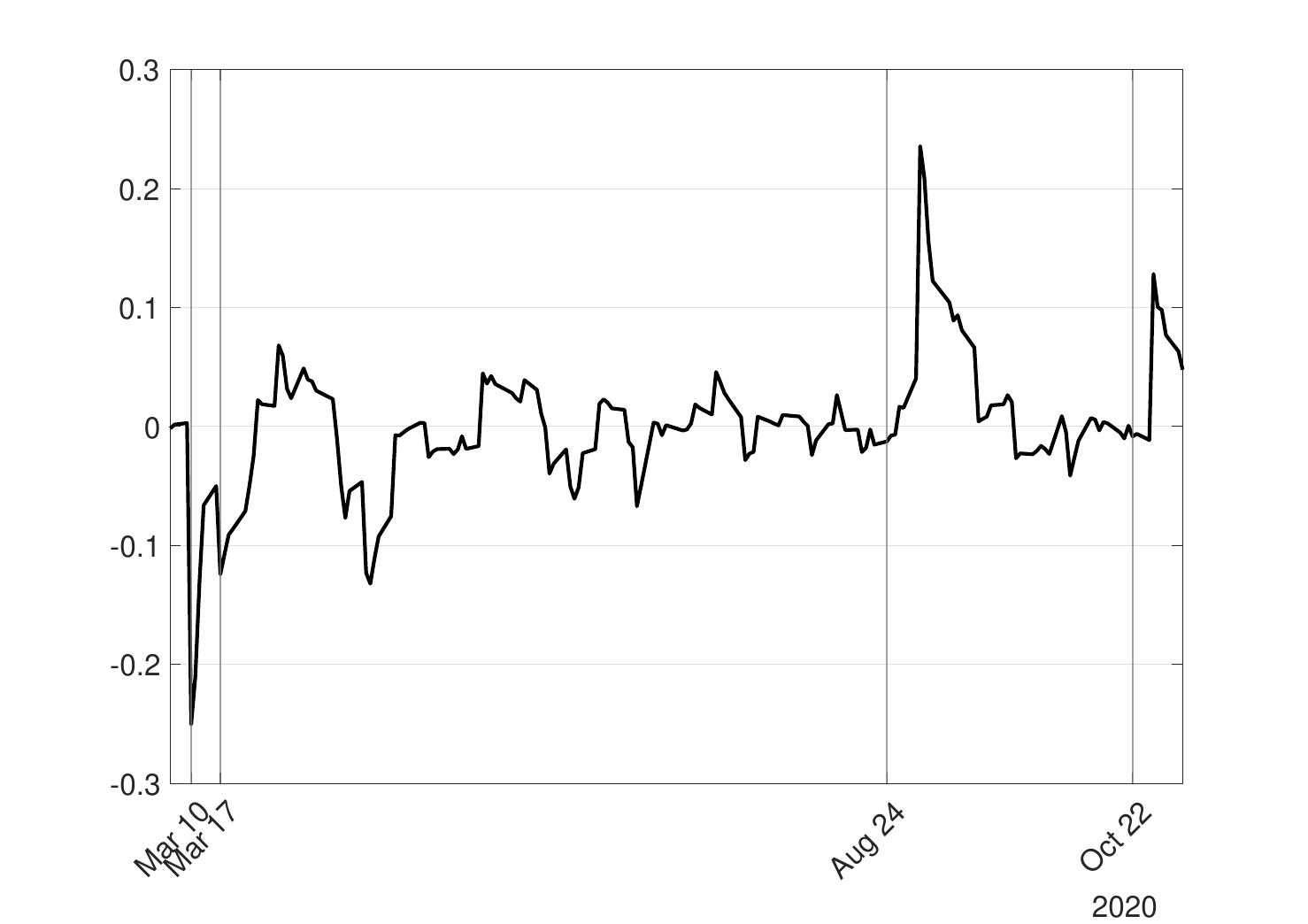}}
	\caption{Estimated conditional correlations (black line) from the DCC model and relevant electoral events (gray vertical lines, see text). Sample period: March 5, 2020 -  November 3, 2020.}
	\label{fig:est_2020}
\end{figure}

\begin{figure}[t]
	\centering
	\subfigure[vs ADS]{\includegraphics[height=4.5cm,width=6.7cm]{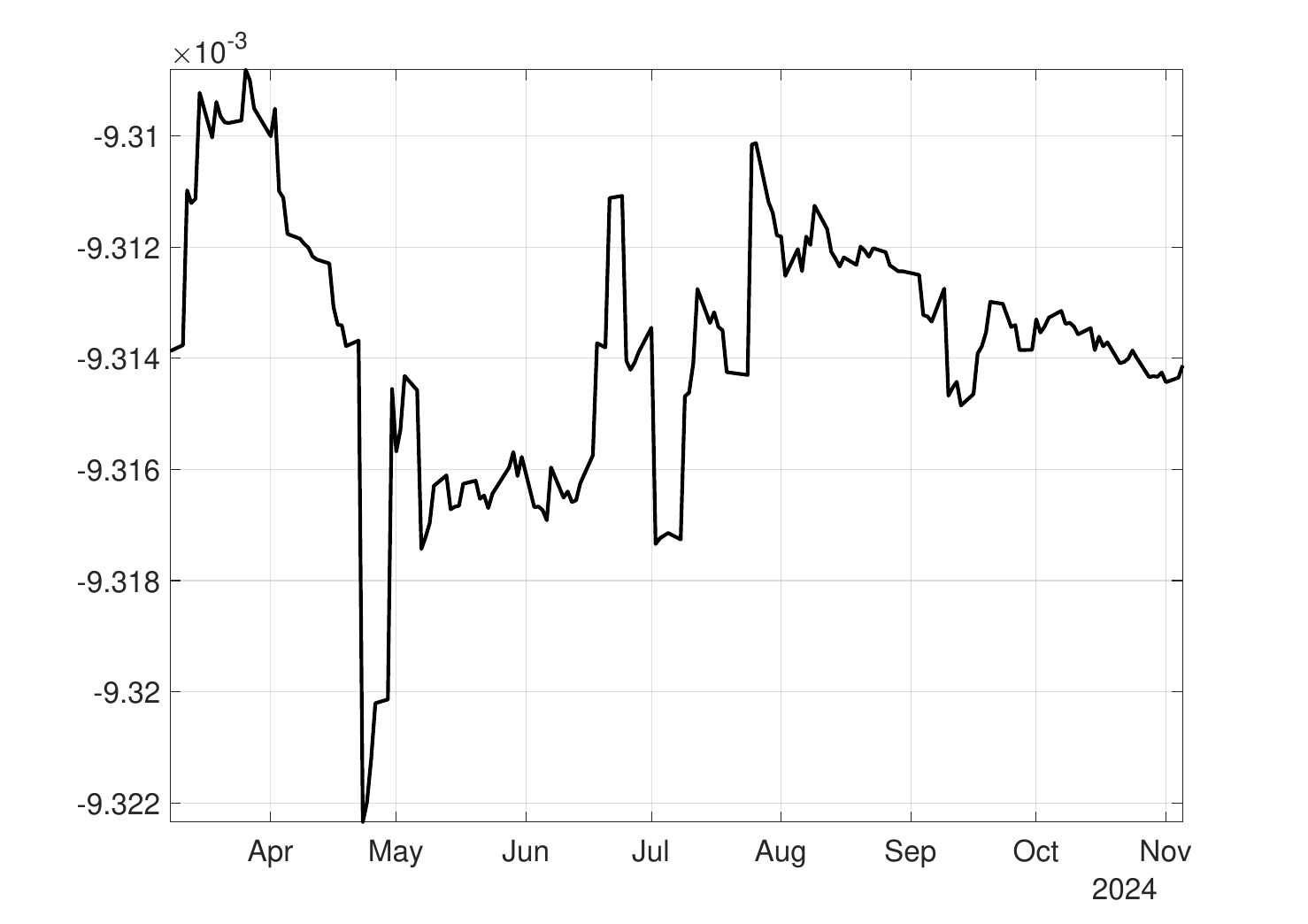}}
	\subfigure[vs inflation]{\includegraphics[height=4.5cm,width=6.7cm]{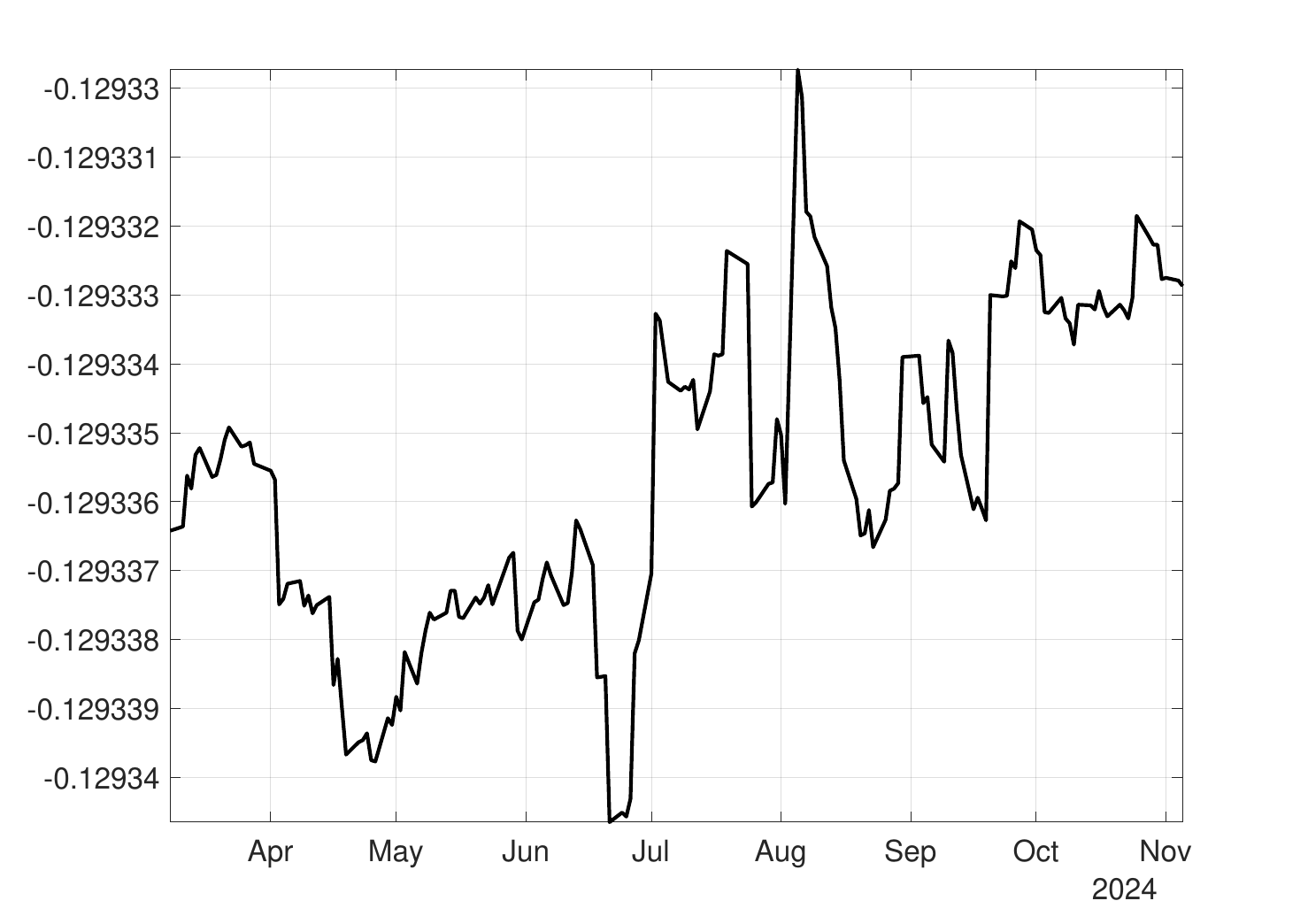}}\\
	\subfigure[vs TPU]{\includegraphics[height=4.5cm,width=6.7cm]{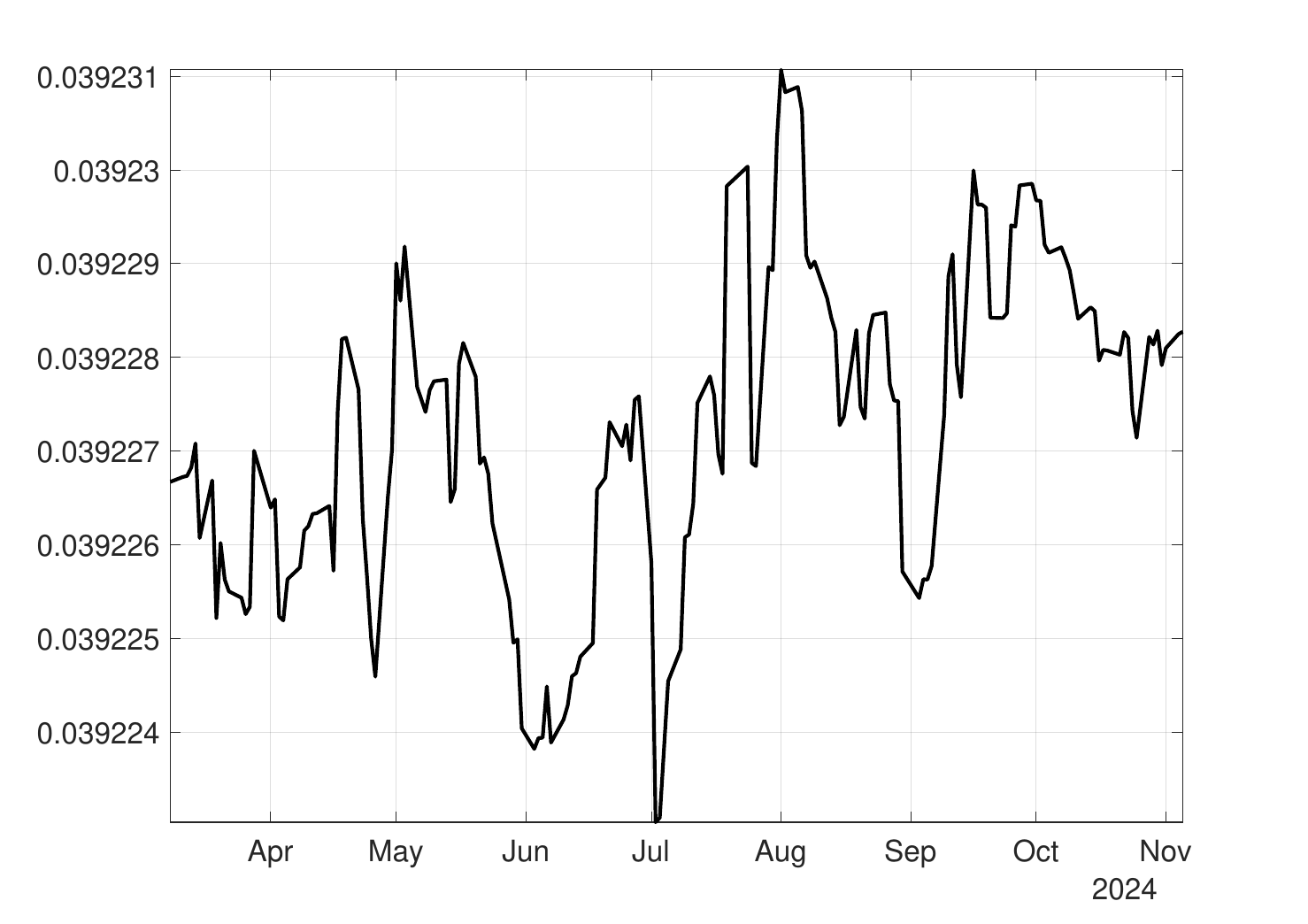}}
	\subfigure[vs VIX]{\includegraphics[height=4.5cm,width=6.7cm]{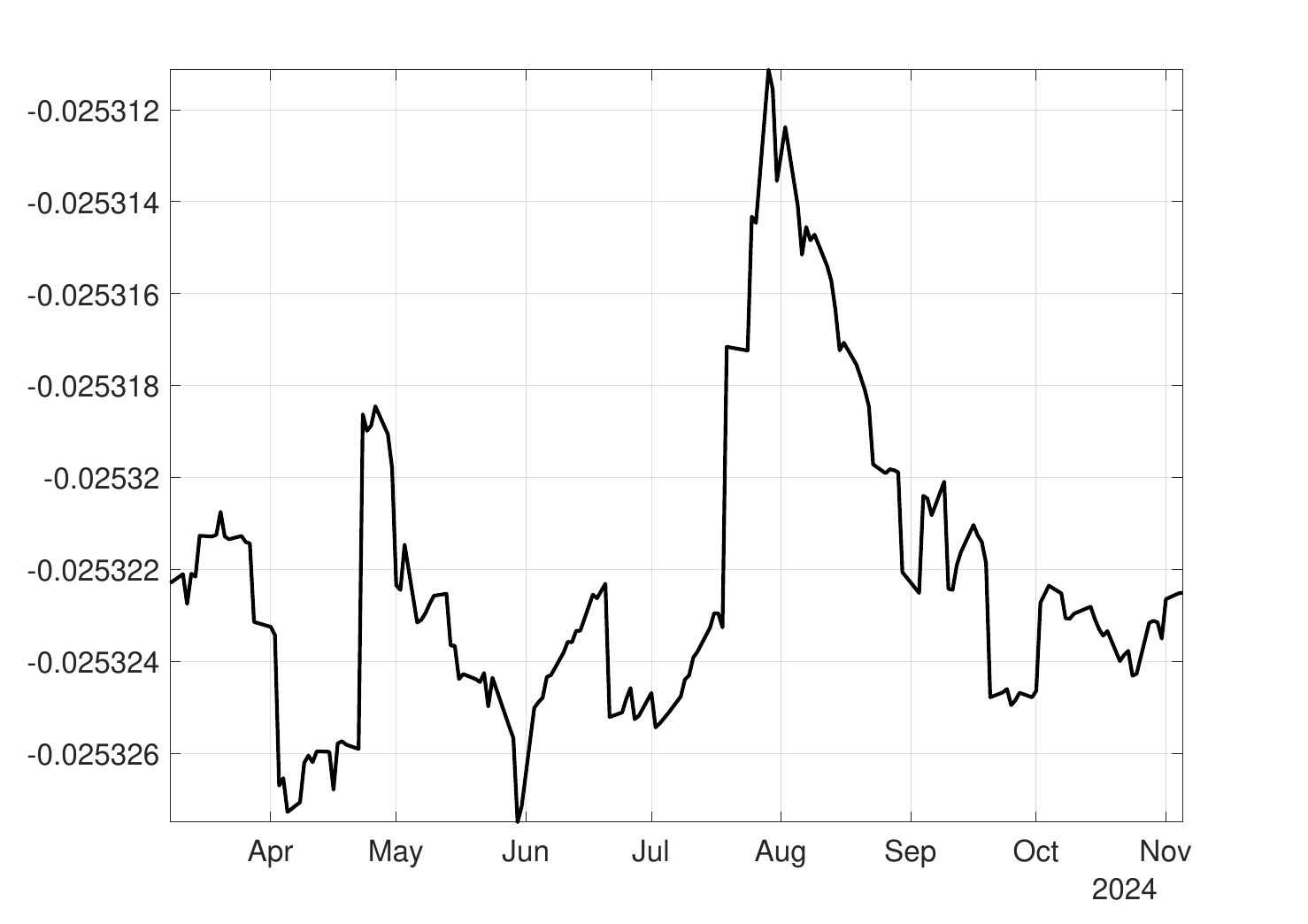}}
	\caption{Estimated conditional correlations (black line) from the DCC model. Sample period: March 8, 2024 -  November 5, 2024.}
	\label{fig:est_2024}
\end{figure}

During the 2020 electoral campaign, the standard DCC appears able to capture the effect of news on correlations, as the estimated coefficient $a=0.032$ is significant at a 10\% level. In this specification, the persistence of correlation is considerable with $b=0.795$. Similarly, in the NLARC model the persistence of correlations remain almost unchanged, while $a$ is not significant. In both specifications, $b$ is not close to 1, meaning that there is room for new information to affect correlation dynamics. The coefficient $\phi_A$ is not statistically significant, so that there is not evidence for different and time-varying coefficients $a_t$ and NLARC could be considered equivalent to the DCC. This is  confirmed by the Likelihood Ratio (LR) test, in panel b),  which does not exceed the $\chi^2$ critical value at the 10\% significance level with one degree of freedom (which correspond to the number of restriction needed to obtain the DCC starting from the NLARC equation). In this regards, the DCC outperforms the CCC, as the LR test statistics ($5.320$) is higher than the $\chi^2_{0.1,2}$ critical value. Coherently, for this period, the DCC is the best model according to the Akaike's Information Criterion, AIC.
A different pictures emerges for the 2024. Both in the DCC and NLARC, $a=0.000$, while $b$ approaches unity. This implies a nearly constant value of the correlations and, for NLARC, an unidentified $\phi_A$ parameter. This result, together with the non-rejection of the LR test both for the DCC vs CCC and DCC vs NLARC, suggests that while the dynamic models possessed the potential for strong time-varying effects, this potential was not activated by the specific events within the 2024 sample period (cf. to Figures \ref{fig:est_2020} and \ref{fig:est_2024} for a visual inspection). Not surprisingly, here the lowest AIC is recorded for the simplest CCC, which is the preferred model. 

\begin{table}[t]
	\centering
	\footnotesize
	\caption{Estimated coefficients from the DCC and NLARC models with robust standard errors \citep{White:1980} in parenthesis. $\chi^2_{0.1,dof}$ refers to the critical value from a $\chi^2$ distribution with 1 and 2 degree of freedom (dof), respectively. }\label{tab:results}
		\begin{tabular}{lccccccc}
			\toprule
			\multicolumn{8}{c}{Estimation results}\\
			                 & \multicolumn{3}{c}{2020}   & & \multicolumn{3}{c}{2024}   \\

                & CCC       & DCC       & NLARC     &  & CCC       & DCC       & NLARC     \\
$\phi_A$          &           &           & 0.150    &  &           &           & 8.063    \\
&           &           & (0.316)  &  &           &           & (15.335)  \\
a                 &           & 0.032    & 0.040    &  &           & 0.000    & 0.000    \\
&           & (0.018)  & (0.025)  &  &           & (0.000)  & (0.000)  \\
b                 &           & 0.795    & 0.793    &  &           & 0.968    & 0.995    \\
&           & (0.055)  & (0.056)  &  &           & (0.006)  & (0.040)  \\
Log-likelihood    & -409.454 & -406.795 & -406.725 &  & -411.807 & -411.807 & -411.807 \\
AIC               & 4.935    & 4.927    & 4.938    &  & 4.962    & 4.986    & 4.998    \\
\midrule

\multicolumn{8}{c}{Diagnostic}\\
$LR_{DCC~vs~CCC}$        & \multicolumn{3}{c}{5.320}        &  & \multicolumn{3}{c}{0.000}        \\
$\chi^2_{0.1,2}$ & \multicolumn{3}{c}{(4.605)}              &  & \multicolumn{3}{c}{(4.605)}              \\
$LR_{DCC~vs~NLARC}$      & \multicolumn{3}{c}{0.140}        &  & \multicolumn{3}{c}{0.000}        \\
$\chi^2_{0.1,1}$ & \multicolumn{3}{c}{(2.706)}              &  & \multicolumn{3}{c}{(2.706)}              \\
			\bottomrule
		\end{tabular}
\end{table}

Figure \ref{fig:est_2020} illustrates the temporal evolution of estimated conditional correlations (from the DCC model) between Trump's polling support and the four indicators considered: the Business Conditions Index (panel a, black line), inflation (panel b), Trade Policy Uncertainty (TPU, panel c), and the VIX (panel d), where major political events  during the 2020 U.S. electoral campaign are reported as vertical gray lines. We detect how specific events had an impact across multiple correlations, notably the primary elections on March 10 (Super Tuesday) and March 17, as well as the Presidential debate on October 22.

In particular, the pronounced decline in correlations observed on Super Tuesday can be attributed to the combined influence of exogenous shocks and electoral dynamics. Exogenous factors, including the COVID-19 pandemic, the collapse in oil prices, and elevated financial market volatility, disrupted the standard relationships between economic fundamentals and political support, as economic conditions deteriorated largely independently of presidential policies. At the same time, Biden's strong performance in the March 10 Democratic primaries reshaped the electoral landscape, increasing the extent to which Trump's polling outcomes reflected perceptions about how he managed the crisis and the positive attitude toward the outcome of the Democratic Party competiton rather than current economic conditions.

Yet, following the March 17 primaries in Florida, Illinois, and Arizona, the correlation between Trump's polls and the Business Conditions Index declined from approximately 0.35 to 0.10, indicating that the sharp deterioration in economic fundamentals -- driven by pandemic-related shutdowns and higher uncertainty -- was no longer directly mirrored in political support. Similarly, the correlation with inflation decreased from around 0.37 to 0.20, suggesting that immediate concerns over unemployment and income losses exceeded the evolution of prices in shaping voters' preferences. In contrast, correlations with both VIX and TPU increased, reflecting heightened sensitivity to financial market volatility and perceived policy uncertainty. The clarification of the Democratic nomination following Biden's victories shifted voter focus toward Trump's crisis management and leadership. Simultaneously, escalating market turbulence amplified perceptions of systemic risk, jointly explaining the strengthened linkage between polls, market volatility, and policy uncertainty during this period.
Finally, after the presidential debate on October 22, 2020, correlations between Trump's poll support and key economic indicators exhibited substantial changes. The correlation with the Business Conditions Index turned negative, while correlations with inflation and VIX declined by approximately 15\% and 25\% respectively, indicating a temporary shift in public attention from underlying economic fundamentals to candidates' debate performance. In other words, following the debate, market participants increasingly treated polling data as a reliable indicator of electoral outcomes. Reduced uncertainty regarding the election amplified the alignment between political perception (polls) and market risk (VIX), with decreasing electoral uncertainty corresponding to lower expected market volatility and vice versa, as resulting from the increasing and positive correlation in the last week of the electoral campaign.

On June 15, 2020, following Biden's attainment of the delegate threshold for the Democratic nomination, the correlation between polls and inflation declined sharply. With the primary effectively concluded, voter attention shifted toward broader political and pandemic-related risks, diminishing the immediate relevance of inflation. Nevertheless, a resurgence of COVID-19 cases in the subsequent days renewed public concern about economic conditions, contributing to a subsequent rebound in the correlation between polls and inflation, which rose rapidly to approximately 0.2. Similarly, following the conclusion of the Democratic National Convention at the end of August 2020, correlations between polls and Trade Policy Uncertainty (TPU) increased. With Trump's nomination secured, voter attention shifted from intra-party competition to perceptions of policy direction and political risk. At the same time, ongoing exogenous factors -- mainly related to the COVID-19 pandemic -- heightened concerns about policy and trade risks. Together, these political and exogenous dynamics explain the strengthened link between polling outcomes and TPU during this period.

Finally, the Republican National Convention attracted significant media attention and focused public debate on Trump's leadership and management of the COVID-19 crisis. In August 2020, financial market volatility remained high and economic uncertainty related to the pandemic persisted, with potential implications for future policies. Consequently, polls became more responsive to perceptions of systemic risk and political instability than to individual economic fundamentals, explaining the increase in the correlation between polls and VIX, the temporary rise in the correlation between polls and business conditions as a proxy for perceived economic risk, and the decline in the correlation between polls and inflation, which became less relevant in a context dominated by political attention and market volatility.


\begin{figure}[t]
	\centering
	{\includegraphics[height=9cm,width=16cm]{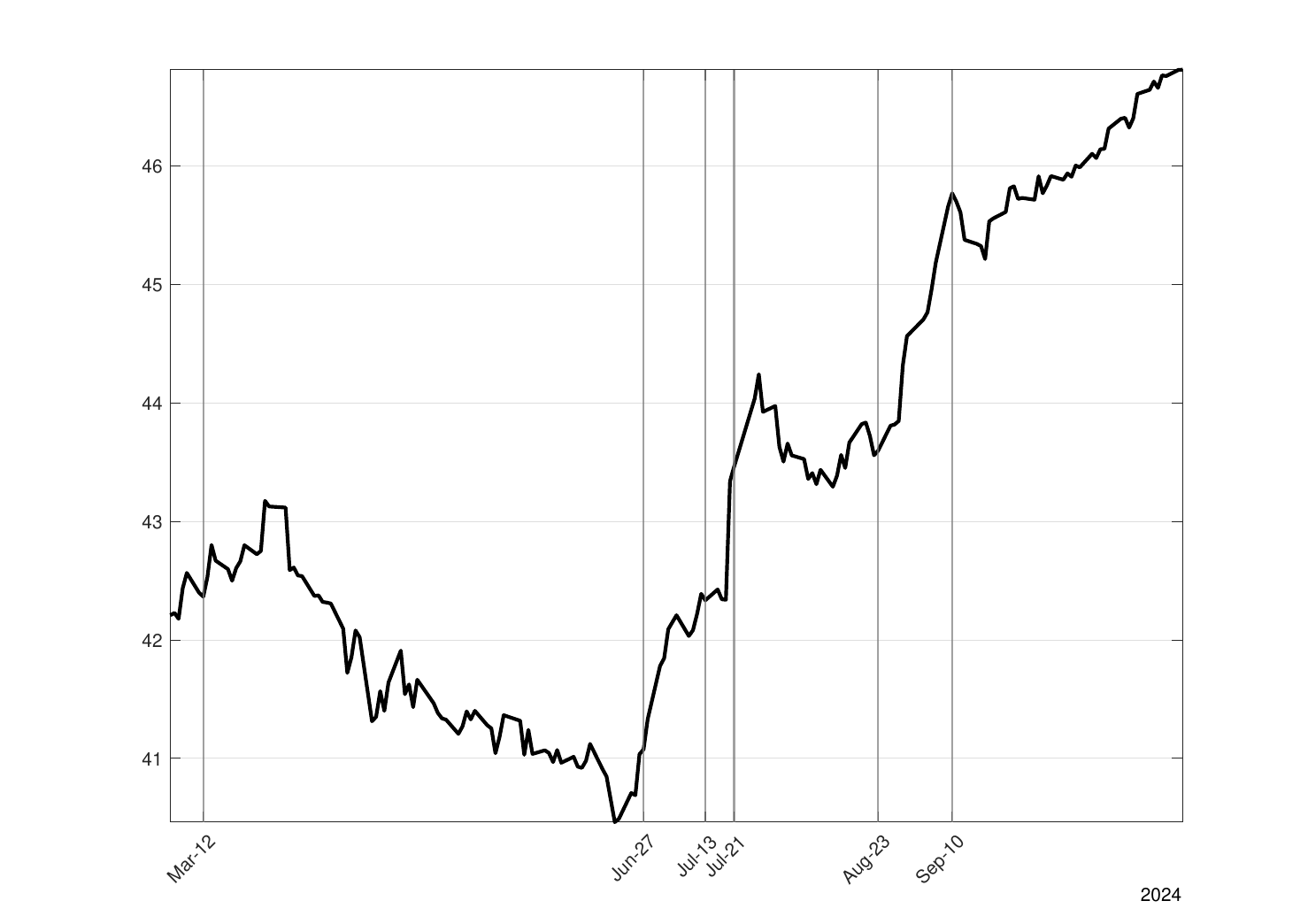}}
	\caption{Time-series of the polls in favor of Trump during the 2024 electoral campaign. The vertical lines represent relevant electoral events (see text).  Sample period: March 8, 2024 -  November 5, 2024}
	\label{fig:polls2024}
\end{figure}

When applying the same methodology to the analysis of the 2024 campaign, the results are surprisingly different. Figure \ref{fig:est_2024} illustrates the dynamics of estimated correlations that show very small fluctuations within a narrow range  around zero, confirming that in the 2024 electoral campaign the preferences for the candidate were less responsive to the evolution of uncertainty. The empirical evidence indicates a larger neutrality to the specific shocks that instead had triggered the reaction to news in the case of the 2020 campaign. We take this change in the profile of correlations to reflect a stronger polarization of opinions less likely to be changed by the evolution of uncertainty about economic conditions. What emerges from our results is that a series of external events during the 2024 election campaign turned out to have a larger impact: the attempted murder on Trump, Biden's announcement and Harris's nomination, Elon Musk's endorsement and Robert F. Kennedy Jr.'s withdrawal from the campaign. The evolution is shown in Figure \ref{fig:polls2024} with vertical lines in correspondence with some events. A first notable event occurs on March 12, when Trump secures the Republican nomination, marking the formal start of his general election campaign. On June 27, the first presidential debate between Biden and Trump takes place, after which Trump's polls show signs of strengthening. The July 13 assassination attempt on Trump, together with Elon Musk's endorsement on the same day, coincides with a further upward shift in support. Shortly thereafter, on July 21, President Biden withdraws from the race and the initial interest in the Harris-Waltz ticket; however, this critical turning point does not appear to reverse the trend in Trump's favor. On August 23, the withdrawal of RFK Jr. removes a third-party contender from the race, consolidating the field. Finally, the September 10 presidential debate marks another focal point, occurring against the backdrop of sustained gains in Trump’s polling advantage.


\section{Final remarks}\label{sec:conclusion}
This paper examined the relationship between electoral polls and different sources of uncertainty, separately during the 2020 and 2024 U.S. presidential elections, focusing on economic, financial, and trade-related indicators. By employing Conditional Correlation models, we documented striking differences between the two electoral periods. In 2020, correlations between Trump’s polling support and the uncertainty measures were clearly time-varying and highly sensitive to both exogenous shocks and political events, with heterogeneous effects across business conditions, inflation, financial volatility, and trade policy uncertainty. By contrast, during the 2024 campaign, correlations were close to zero, stable, and largely unresponsive to news, indicating a weaker interplay between electoral dynamics and macroeconomic uncertainty.

These findings highlight the importance of considering the broader economic and political environment when studying electoral polls, as correlations may change substantially across periods characterized by different types and magnitudes of shocks. The evidence also suggests that market expectations and voter sentiment may interact more strongly in contexts of heightened uncertainty, while such links can vanish in calmer environments or when external and exceptional events (see, for example,  the attack on Trump, Harris's candidacy, Elon Musk's endorsement) directly influence the electoral campaign regardless of economic uncertainty. 

Given that the DCC models are typically used for financial time series involving thousands of observations, this analysis may be affected by the short duration of the two intervals. It might be interesting to replicate the analysis by considering longer time series related to political dynamics, such as presidential approvals, spanning an entire administrative cycle, especially in light of the strong uncertainty induced by some of Trump's announcements in the first few months of his second mandate.

Furthermore, future research could extend this analysis by considering alternative measures of uncertainty and applying nonlinear models that can capture regime shifts more explicitly; for example, by adopting Markov Switching or Smooth Transition models for the variance part \citep{Gallo:Otranto:2015,Gallo:Otranto:2018} and nonlinear models for the correlation part, as described in \citet{Bauwens:Otranto:2016}.

\subsection*{Disclosure statement}

No competing interest is declared.

\subsection*{Data Availability Statement}
Data are available upon request.

\bibliography{biblio.bib}

\end{document}